\documentclass[iop]{emulateapj}
\usepackage{apjfonts,amsmath}
\usepackage{color}

\newcommand{\black}{\color[rgb]{0,0,0}}
\newcommand{\gal}{B0218$+$357}
\newcommand{\form}{\mbox{H$_2$CO}}

\newcommand{\kms}{\mbox{~km~s$^{-1}$}}
\newcommand{\kmspc}{\mbox{~km~s$^{-1}$~pc$^{-1}$}}
\newcommand{\vgrad}{\mbox{(km~s$^{-1}$~pc$^{-1})^{-1}$}}
\newcommand{\pcc}{~\mbox{cm$^{-3}$}}
\newcommand{\psc}{~\mbox{cm$^{-2}$}}
\newcommand{\nh}{\mbox{$n$({\rm H}$_2$)}}
\newcommand{\Nh}{\mbox{$N$({\rm H}$_2${\rm CO})}}
\newcommand{\oNh}{\mbox{$N$(o-H$_2$CO)}}
\newcommand{\cmone}{\mbox{$1_{10}$$-$$1_{11}$}}
\newcommand{\cmtwo}{\mbox{$2_{11}$$-$$2_{12}$}}
\newcommand{\tcmb}{\mbox{T$_{\rm CMB}$}}
\newcommand{\tkin}{\mbox{T$_{\rm kin}$}}
\newcommand{\tx}{\mbox{T$_{{\rm ex}}$}}
\newcommand{\op}{\mbox{o-\form/p-\form}}
\newcommand{\ten}[1]{\mbox{$10^{#1}$}}

\shorttitle{Formaldehyde Anti-Inversion in B0218$+$357}
\shortauthors{Zeiger \& Darling}

\begin{document}
\slugcomment{\noindent The Astrophysical Journal {\bf 709} (2010) 386}

\title{Formaldehyde Anti-Inversion at $z=0.68$ in the Gravitational Lens B0218$+$357}

\author{\black Benjamin Zeiger\altaffilmark{1,2} \&  Jeremy Darling\altaffilmark{1,3,4}}
\altaffiltext{1}{Center for Astrophysics and Space Astronomy,
Department of Astrophysical and Planetary Sciences,
University of Colorado, 389 UCB, Boulder, CO 80309-0389, USA}
\altaffiltext{2}{benjamin.zeiger@colorado.edu}
\altaffiltext{3}{NASA Lunar Science Institute, NASA Ames Research Center, Moffett Field, CA, USA}
\altaffiltext{4}{jdarling@colorado.edu}

\begin{abstract}

We report new observations of the \cmone\ (6~cm) and \cmtwo\ (2~cm) transitions of ortho-formaldehyde (o-\form) in absorption at $z=0.68466$ toward the gravitational lens system \gal.  Radiative transfer modeling indicates that both transitions are anti-inverted relative to the 4.6~K cosmic microwave background (CMB), regardless of the source covering factor, with excitation temperatures of $\sim$1~K and $1.5-2$~K for the \cmone\ and \cmtwo\ lines, respectively.  Using these observations and a large velocity gradient radiative transfer model that assumes a gradient of 1~\kmspc, we obtain a molecular hydrogen number density of $2\times\ten{3}\pcc<\nh<1\times\ten{4}\pcc$ and a column density of  $2.5\times10^{13}\psc<\oNh<8.9\times10^{13}\psc$
, where the allowed ranges conservatively include the range of possible source covering factors in both lines.  The measurements suggest that \form\ excitation in the absorbing clouds in the \gal\ lens is analogous to that in Galactic molecular clouds: it would show \form\ absorption against the CMB if it were not illuminated by the background quasar or if it were viewed from another direction.

\end{abstract}



\section{Introduction}
Formaldehyde (\form) centimeter transitions have been observed with excitation temperatures less than the temperature of the cosmic microwave background ($\tx<\tcmb$) in the Milky Way \citep{palmer69}.  The ``anti-inversion,'' or non-thermal absorption, is pumped by collisions, coupling \form\ excitation to the number density \nh\ of molecular hydrogen even at kinetic temperatures as low as $\sim$10~K \citep{tc69,evans75,glmg75}. Thus, \form\ provides a density and temperature diagnostic for molecular gas \citep{muhle,mangum08}, and it often produces detectable lines before dense molecular clouds begin to form stars and warm sufficiently to emit lines from the usual molecular tracers such as CO and HCO$^+$. 

\begin{figure}
\epsscale{1.4}
\plotone{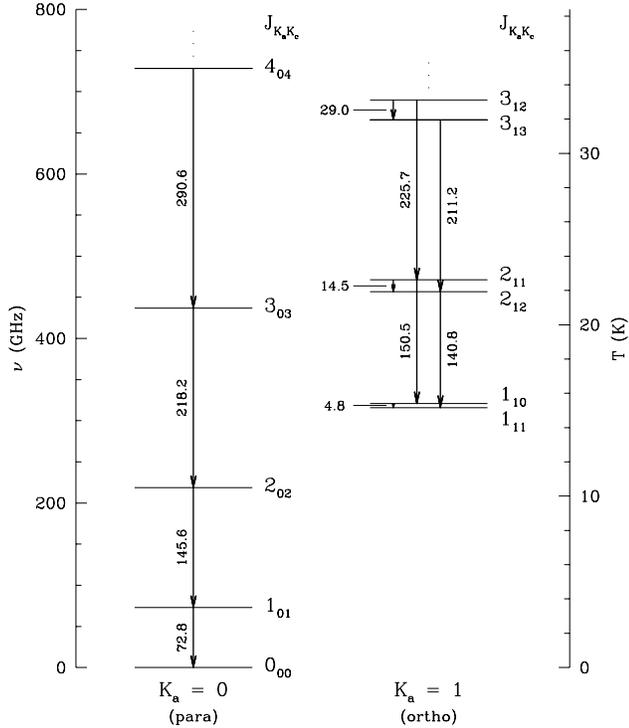}
\caption{Formaldehyde (H$_2$CO) energy level diagram showing only the lower levels and $K_a<2$.  The indicated transitions are in GHz.\label{H2CO_levels}}
\end{figure}

The rotation states of \form\ are defined by the three quantum numbers $J_{K_aK_c}$ (total angular momentum and the angular momenta about the axes with the smallest and largest moments of inertia, respectively).  The lower energy transitions for $K_a<2$ are shown in Figure~\ref{H2CO_levels}.  The spin symmetry of the hydrogen nuclei 
defines the ortho ($K_a=$ odd) and para ($K_a=$ even) states.  Since the spin symmetry is about the $a$ axis, which is aligned with the electric dipole along the C--O double bond and thus unchanged by dipole interactions, the mean transition time from o-\form\ to p-\form\ states (or vice versa) is longer than the mean life of the molecule, and the ortho/para ratio preserves information about the dominant formation channel: a ratio of unity indicates catalysis on the surface of cold ($\tkin\lesssim15$~K) dust grains, while a ratio of 3 indicates gas-phase formation \citep{kahane,dick99}.

The most commonly detected o-H$_2$CO lines are the $\Delta J=0$, $\Delta K_c=\pm1$ ``$K$-doublet'' $1_{10}-1_{11}$ 4.8~GHz (6~cm) and  $2_{11}-2_{12}$ 14.5~GHz (2~cm) centimeter transitions  and the $\Delta J=\pm1$, $\Delta K_c=\pm1$ millimeter transitions connecting the $J=1$ states to the $J=2$ states at 140.8 and 150.5 GHz  ($2_{12}$$-$$1_{11}$ and $2_{11}$$-$$1_{10}$, respectively). The coupling of the centimeter and millimeter transitions allows the $K$-doublet transitions to be driven significantly out of local thermodynamic equilibrium (LTE) and makes \form\ an {\it in situ} probe of conditions in molecular clouds.  The relative excitation of $K$-doublets is a powerful densitometer \citep{mangum08}, while the relative  excitation of millimeter lines is a molecular thermometer \citep{muhle}. The $1_{10}-1_{11}$ and \cmtwo\ transitions have both been observed with population anti-inversions, causing stimulated absorption of the CMB, and the \cmone\ transition has been observed as a maser \citep{hp74,evans75,araya08}.
\begin{deluxetable*}{ccccrrlcc}[t]
\tabletypesize{\footnotesize}
\tablewidth{0pt}
\tablecolumns{9}
\tablecaption{Journal of o-\form\ Observations Toward \gal\label{tab:journal}}
\tablehead{
\colhead{Telescope} & 
\colhead{Receiver} & 
\colhead{UT Date} & 
\colhead{Transition} &
\colhead{$\nu_\circ$}&
\colhead{$\nu_{{\rm obs}}$} &
\colhead{$t_{int}$} & 
\colhead{$\Delta$v\,\tablenotemark{a}} &
\colhead{rms}\\
\colhead{} & 
\colhead{} & 
\colhead{} & 
\colhead{} & 
\colhead{(GHz)} & 
\colhead{(GHz)} & 
\colhead{(s)} & 
\colhead{(km s$^{-1}$)} &
\colhead{(mJy)}
}
\startdata
Arecibo	     & S-low  & 2004 Aug 3--4 & $1_{10}$$-$$1_{11}$ & 4.829660(1) & 2.8668455(6) & 3600\tablenotemark{b} & 1.3 & 0.8 \\
GBT & X & 2006 Apr 26 & $2_{11}$$-$$2_{12}$ & 14.488479(1) & 8.6002392(6) & 3280 & 2.1 & 1.1
\enddata
\tablenotetext{a}{$\Delta$v is the rest-frame spectral velocity resolution of the final, smoothed spectrum.}
\tablenotetext{b}{Two 1800 s sessions from consecutive days were combined for the final spectrum.}
\end{deluxetable*}

We report observations of the \cmone\ and \cmtwo\ transitions in absorption toward \gal, a gravitational lens system with two images separated by 0\farcs33. The lens lies at a redshift of $z=0.68466(4)$ \citep{browne93}, while the background quasar is at $z=0.944(2)$ \citep{cohen03}.  At low frequencies, the ``gravitational lens with the smallest separation'' of its images also displays the smallest known Einstein ring, with a diameter of 0\farcs335 \citep{pat93}.  The two images, dubbed A and B in order of radio brightness, display significantly different absorption characteristics, with B unobscured while A shows strong molecular absorption lines and is heavily obscured by dust in optical observations \citep{wik95, grundahl95}.  Both A and B show the background source to have a core-jet morphology.

Many molecules have been observed at the redshift of the lens in absorption against the strong continuum of image A, including CO, HCO$^+$, HCN, H$_2$O, NH$_3$, OH, and \form; typical line widths are 10$-$15\kms\ \citep{wik95, combes97, henkel05,kanekar03,menten96}. \ion{H}{1} observations indicate the lensing galaxy to be gas-rich, with a column density $N(\mbox{\ion{H}{1}}) = 4 \times\ten{18}(T_s/f)$\psc\ across a line width (FWHM) of 43\kms, where $T_s$ is the spin temperature and $f$ is the \ion{H}{1} covering factor \citep{carilli93}. Four spectrally resolved absorption components can be identified in front of image A in high-resolution interferometric HCO$^+$ observations \citep{muller}. 

\form\ was first observed in \gal\ by \citet{menten96} in the \cmtwo\ transition in absorption.  \citet{jethava07} followed these observations with a six-line study of \form\ in \gal, adding observations of the lowest two $\Delta J=1$ rotational transitions of both the ortho and the para species, as well as the \cmone\ transition of o-\form.  Using the 55~K kinetic temperature derived from NH$_3$ observations \citep{henkel05} and a large velocity gradient (LVG) model to determine the non-thermal \form\ excitation due to collisions with H$_2$, \citet{jethava07} determined a spatial number density of $\nh<\ten{3}$\pcc\ and a column density of $\Nh\approx5.2\times\ten{13}$\psc\ (no confidence interval is reported) in the strongest absorption component, with a weaker component adding $\sim$15\% to the column density.  They measured an \op\ ratio of $2.0-3.0$ with a best fit of 2.8, giving $\oNh=3.8\times\ten{13}$\psc.  They do not address the issue of \form\ excitation temperatures, which are of interest because anti-inversion of the lines would imply that the gas in the \gal\ lens would be visible in absorption against the CMB even without a background quasar to illuminate it.  Obtaining excitation temperatures of \form\ lines in \gal\ will help to set \form\ observability conditions in non-lensing (or otherwise unilluminated) galaxies.

The \citet{mangum08} survey of local galaxies demonstrates the feasibility of determining both \nh\ and \oNh\ strictly from the \cmone\ and \cmtwo\ transitions.  The relative strengths of the transitions are set by the degree of collisional excitation, giving a precise densitometer that, in turn, yields the excitation temperatures of the transitions.  In our analysis, we use the two-line densitometry method with new data, including \cmone\ observations of higher spectral resolution and signal to noise ratio than that obtained by \citet{jethava07} and \cmtwo\ observations with lower signal to noise (4.7 versus 8.8) but comparable resolution to those of \citet{menten96}. Our analysis, while also yielding \oNh\ and \nh\ from an exhaustive search of the density space with our LVG model, focuses on line excitation temperatures in \gal\ to show that the o-\form\ observed in absorption is anti-inverted relative to the 4.6~K \tcmb\ at $z=0.68$, making it analogous to Galactic molecular clouds that show anti-inversion of the $K$-doublet transitions and thus CMB absorption. 

In Section \ref{sec:obs}, we describe our observational and data reduction procedures used to acquire and reduce the new high resolution spectra of the \cmone\ and \cmtwo\ o-\form\ transitions in \gal.  The properties of both lines are reported in Section \ref{sec:results}, 
and we use those results to determine \nh, \oNh, and line excitation temperatures in Section \ref{sec:analysis}. We show that the absorbing clouds in \gal\ are analogous to Galactic molecular clouds that show anti-inversion of o-\form\ centimeter lines relative to the $z=0$ \tcmb, and that the anti-inversion of $K$-doublet transitions persists up to the $6_{15}-6_{16}$ (101~GHz rest) transition (Section \ref{sec:discussion}).  The consequences of our results are discussed in Section \ref{sec:conclusions}.

\begin{deluxetable*}{lcccrrcl}[t]
\tablewidth{0pt}
\tablecolumns{8}
\tablecaption{Measured and Calculated Formaldehyde Line Properties}
\tablehead{
\colhead{Transition} & 
\colhead{$\nu_\circ$} &  
\colhead{Compt} & 
\colhead{${\rm v}_{{\rm obs}}-{\rm v}_{{\rm sys}}$} &
\colhead{Depth} &
\colhead{FWHM} &
\colhead{$\tau_{{\rm app}}$} &
\colhead{$\int \tau_{{\rm app}}\,dv$} 
\\
\colhead{} &
\colhead{(GHz)} &  
\colhead{} &
\colhead{(km s$^{-1}$)} &
\colhead{(mJy)} &
\colhead{(km s$^{-1}$)} &
\colhead{} &
\colhead{(km s$^{-1}$)} 
}
\startdata
$1_{10}$$-$$1_{11}$ &  4.829660(1) &  1 & +3.7(0.4)  & 9.6(0.4) & 12.3(0.6) & 	     &	   \\ 
			     &   & 2 &$-$9.8(2.4) & 1.8(0.3) & 14.2(4.0) &  	     &	   \\ 
			     & &Total& +3.4(0.4)  &10.8(0.8) & 12.6(0.6) & 0.017(1) & 0.239(8) \\ 
\\
$2_{11}$$-$$2_{12}$ & 14.488479(1)&  1 & +4.2(0.8)  & 5.2(1.1) &  8.5(1.5)   & 0.008(2)& 0.083(10)
\enddata
\tablecomments{${\rm v}_{{\rm obs}}-{\rm v}_{{\rm sys}}$ is the {\it rest-frame} centroid velocity offset from the assumed systemic redshift, $z=0.68466$, FWHM is the rest-frame line width, and $\tau_{{\rm app}}$ is the maximum apparent optical depth in the line, assuming a covering factor of unity and a continuum in image A of $650\pm16$ and $663\pm22$~mJy in the $1_{10}$$-$$1_{11}$ and $2_{11}$$-$$2_{12}$ 
lines, respectively (Section \ref{sec:results}).  
Uncertainties are strictly statistical uncertainties
from the spectra and do not reflect the systematic uncertainties associated
with flux calibration or the radio continuum determinations. Reported quantities for the \cmone\ ``Total'' and \cmtwo\ lines are measured directly from the data and are not dependent on profile fitting; the individual components of the \cmone\ line are from Gaussian fits.}
\label{tab:H2CO}
\end{deluxetable*}

\section{Observations and Data Reduction}\label{sec:obs}

\subsection{Arecibo Telescope: \cmone\ Line}
We observed the \cmone\ 4.829660(1)~GHz transition\footnote{All line frequencies are from the JPL molecular line database \citep{pic98}.} of o-H$_2$CO,
redshifted to 2.8668455(6)~GHz, toward 
B0218$+$357 at the Arecibo radio telescope\footnote{The Arecibo Observatory is part of the National Astronomy and Ionosphere Center, which is operated by Cornell University under a cooperative agreement with the National Science Foundation.} in August 2004 (Table \ref{tab:journal}). Observations were conducted with a 6.25~MHz band centered on the redshifted line, using 5~minute position-switched scans with a calibration diode fired after each position-switched pair 
and spectral records recorded every 6 s.  Total integration time was 3600~s.  The autocorrelation spectrometer used nine-level sampling in two (subsequently averaged) polarizations.  Bandpasses were divided into 1024~channels and Hanning smoothed to 512~channels. Rest-frame velocity resolution was 1.3~km~s$^{-1}$ (see Table \ref{tab:journal}). The observed band was interference-free in the vicinity of the observed line.

Records were individually calibrated and bandpasses were flattened using the calibration diode and the corresponding off-source records. Records and polarizations were subsequently averaged, and a polynomial baseline 
with variations much larger than the line was fit to and subtracted from the spectrum.  Systematic flux calibration errors in these data are of order $10\%$.  All data reduction was performed in the Astronomical Information Processing System$+$$+$, AIPS$+$$+$.\footnote{AIPS$+$$+$ is freely available for use under the Gnu Public License. Further information may be obtained from http://aips2.nrao.edu.} Arecibo spectra often show spectral standing waves due to resonances of the strong continuum flux densities %
within the telescope superstructure. 
The line width was significantly less than the size of the standing wave features, so the spectrum was not significantly affected.

\subsection{Green Bank Telescope: \cmtwo\ Line}
We observed the $2_{11}- 2_{12}$ 14.488479(1)~GHz transition of H$_2$CO, redshifted to 8.6002392(6)~GHz, toward \gal\ with the Green Bank Telescope\footnote{The National Radio Astronomy Observatory is a facility of the National Science Foundation operated under cooperative agreement by Associated Universities, Inc.} (GBT) on 2006 April 26  (Table \ref{tab:journal}). Observations were conducted with a 50~MHz bandpass centered on the redshifted line in 5~minute position-switched scans with spectral data recorded every 3~s and a winking calibration diode firing during every other record. The total on-source integration time was 3280~s. The autocorrelation spectrometer used nine-level sampling in two (subsequently averaged) linear polarizations.  Bandpasses were divided into 16384 channels, Hanning smoothed to 8192 independent channels, and 10-channel Gaussian smoothed to a rest-frame velocity resolution of 2.1~km~s$^{-1}$.  Records were individually calibrated and bandpasses were flattened using the calibration diode and the corresponding off-source records. The individual spectra were interference-free  and fairly flat, but the DC levels of the individual scans showed significant fluctuations of order $10\%$. Scans and polarizations were subsequently averaged, and a fifth-order polynomial baseline was fit to the full 50~MHz bandwidth and subtracted.  Systematic flux calibration errors in these data are of order $10\%$.  All data reduction was performed in GBTIDL.\footnote{GBTIDL (\url{http://gbtidl.nrao.edu/}) is the data reduction package produced by NRAO and written in the IDL language for the reduction of GBT data.}

\begin{figure}
\epsscale{1.30}
\plotone{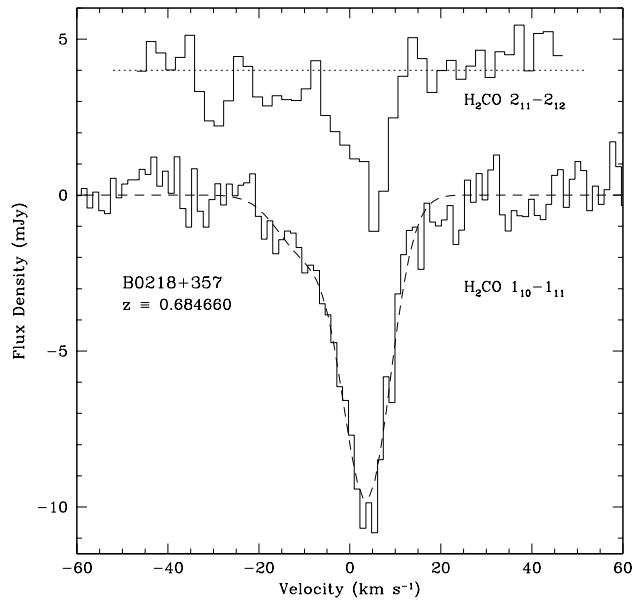}
\caption{Formaldehyde $1_{10}- 1_{11}$ (6~cm) and $2_{11}- 2_{12}$ (2~cm) absorption toward the gravitational lens system B0218$+$357. The rest-frame velocity scale assumes a heliocentric redshift of $z=0.684660$, and the spectral resolutions are 1.3 and 2.1 km s$^{-1}$ in the lower and upper lines, respectively. The dashed line shows a two-component Gaussian fit to the 6~cm line profile. The 2~cm line spectrum, also detected by \citet{menten96}, is offset by 4~mJy, with the zero point indicated by the dotted line.
\label{B0218_H2CO_cm}}
\end{figure}

\section{Results}\label{sec:results}
The Arecibo spectrum of the $1_{01}- 1_{11}$ transition 
requires two Gaussian components for a good fit to the line profile (Figure~\ref{B0218_H2CO_cm}). The rest frame velocity resolution is 1.3\kms. Table~\ref{tab:H2CO} lists the two fit components and the properties of the total line. The total line properties are not based on Gaussian fits; they are computed directly from the spectrum. All velocities are in the rest frame with respect to a heliocentric redshift of $z=0.68466$. The integrated optical depth in this transition is $0.239(8)$ km s$^{-1}$ over $-20\kms\leq v \leq +17$\kms\ assuming a continuum of $650\pm16$~mJy and a unity covering factor (see Sections \ref{sec:analysis-continuum} and \ref{sec:fc}).

The GBT detection of the $2_{11}-2_{12}$ transition has a lower signal to noise ratio than the Arecibo \cmone\ transition.  Despite this and the lower spectral resolution of the GBT spectrum (2.1 versus 1.3\kms), both lines display similar profiles (Figure~\ref{B0218_H2CO_cm}), although the \cmtwo\ detection is insufficient to justify a second profile component as seen in the \cmone\ spectrum and in the data of \citet{jethava07}.  The properties listed in Table \ref{tab:H2CO} for this line have been measured directly from the data.  The integrated optical depth, spanning the same  velocity range as the 4.8~GHz line, is  $0.083(10)$ km s$^{-1}$, assuming a continuum of $663\pm22$~mJy and a  unity covering factor (see Sections \ref{sec:analysis-continuum} \S \ref{sec:fc}).

The observed profiles are narrower than the 15\kms\ observed in CO, HCO$^+$, and HCN \citep{combes96}, but broader than the NH$_3$ lines that range from 6.6(1.6) to 9.3(1.8)\kms\ \citep{henkel05}. OH spectra (1.67~GHz rest) clearly show a two-component profile \citep{kanekar03}, although at 20 and 27\kms\ the line components are broader than the $\sim$10\kms\ \form\ lines. Our detection of the \cmtwo\ transition is comparable to that of \citet{menten96}.  \citet{jethava07} report a much narrower \cmone\ line --- with a FWHM of 5.5(5)\kms\ --- but our high signal to noise ratio and resolution make our measurement more statistically robust.  Both of our observations are coincident in velocity space with previous molecular detections in \gal.

\section{Analysis}\label{sec:analysis}

\subsection{Source Continuum}
\label{sec:analysis-continuum}
In general, while the continuum determinations may be uncertain and lead to errors in the absolute optical depth and total column density, the computation of quantities that depend on ratios of (low) optical depths, such as excitation temperatures and \nh, will be significantly less affected, 
provided that systematic errors are consistent across frequency. 

The continuum emission at 2.9~GHz (the frequency of the redshifted 4.8~GHz \cmone\ transition) is nearly flat as a function of frequency for images A and B 
\citep{odea92,pat93} but steeper for the Einstein ring and the extended emission halo. At 5~GHz, the ratio of the two compact flat spectrum components A:B is $\sim3.0$ and the Einstein ring emission accounts for 10\%--$20\%$ of the total emission \citep{odea92,pat93}.  The two sources show time variability with a 10.5~day period, but the variability is constrained to be $<8$\%\ at low frequencies \citep{mittal}.  Extended emission is detected at 1.465 and 1.63~GHz at the $\sim10\%$ level, and additional structure is seen at 5~GHz \citep{odea92,pat93}.

Comparison to other molecules confirms the association of \form\ absorption with image A \citep{menten96,muller}. The appropriate continuum level to use in apparent optical depth calculations for the \cmone\ transition is thus the total continuum at 2.9 GHz (8.6~GHz for the \cmtwo\ line) less image B, the Einstein ring emission, and any extended source structure.  Interpolating the flux densities for images A and B and the extended emission from interferometric observations allows measurement of the fraction of the continuum from image A, a fraction crucial to disentangling the absorption in front of image A from the unabsorbed continuum of image B and other extended structure.

The flux density of image A at 2.9 GHz can be interpolated from a linear fit (in the log) to the flux density versus frequency of image A. 
We use the very long baseline interferometry (VLBI) observations of \citet{mittal} because these straddle 2.9~GHz and because these observations, of all the available interferometric measurements, were taken closest in time to our spectra (2002 January versus 2004 August for the Arecibo \cmone\ observations and 2006 April for the GBT \cmtwo\ observations).   The continuum, interpolated from a fit to the image A data points spanning 2.25--15.35~GHz (the 1.65~GHz datum is dropped due to a steep downturn below 2.25~GHz; see \citealt{mittal}), is $650\pm16$~mJy, which neglects systematic and calibration errors typically of order $10\%$.  The image A continuum is nearly flat, and interpolation error is less of a concern than overall flux calibration.  \citet{mittal} degrade their resolution to $50\times50$~mas, comparable in resolution to MERLIN observations by \citet{pat93} to account for the loss of flux in VLBI observations, yielding similar flux densities to within $10\%$.  Note that extrapolations from lower resolution (and much older) observations predict higher flux densities at 2.9~GHz: \citet{odea92} predict $868\pm64$ mJy, and \citet{pat93} predict $791\pm68$ mJy.  
We adopt $S_{\rm A}(2.867\mbox{ GHz}) = 650\pm16$~mJy, bearing in mind that this estimate could potentially have large systematic errors.
For the 14~GHz line redshifted to 8.6~GHz, we use the same VLBI data and continuum estimation method, adopting $S_{\rm A}(8.600\mbox{ GHz}) = 663\pm22$~mJy. The VLBI observations of \citet{mittal} are degraded to 10~mas at 8.4~GHz and 5~mas at 15~GHz. 

\subsection{Source Covering Factor}
\label{sec:fc}
Calculation of the true optical depth of each line requires knowledge of the covering factor $f_\nu$ describing the fraction of image A obscured by the absorbing cloud at the line frequency $\nu$. The covering factor corrects the observed line intensity to account for dilution of the line due to incomplete coverage of the continuum source by the absorbing cloud.  Thus, the optical depth $\tau_{{\rm line}}$ of the line takes the form \begin{equation}\label{eqn:tau}
  \tau_{{\rm line}} = -\ln\left( 1-\frac{S_{{\rm line}}}{f_\nu S_\nu} \right),
\end{equation}
where $S_{{\rm line}}$ and $S_\nu$ are the line and continuum fluxes, respectively. Millimeter observations at $\nu>100$~GHz show $f_\nu\gtrsim0.77$ \citep{wik95}, consistent with a covering factor of unity in front of image A and 0 in front of image B; the flux ratio (A/B) is close to 4 at 100~GHz \citep{muller}.

Due to the increased solid angle of image A at lower frequencies \citep{mittal}, the covering factors for the \cmone\ and \cmtwo\ transitions could be significantly less than unity even after excluding flux from image B and extended emission.  \citet{kanekar03} estimate the covering factor to be 0.4 at 990~MHz (redshifted 18~cm OH), and \citet{jethava07} estimate the covering factor to be of the form $f_\nu=(\nu/100)^{x}$ for $\nu$ in GHz, determining that $0.0<x<0.5$ by interpolating between the millimeter observations of \citet{wik95} and the 990~MHz OH observations of \citet{kanekar03}. (Note that, in arriving at this range, they refer to the OH lines as being at 850~MHz, the redshifted frequency of the \ion{H}{1} line.)  For discussion, \citet{jethava07} adopt the most extreme frequency dependence of $x=0.5$, which gives covering factors of 0.17 and 0.29 for the \cmone\ and \cmtwo\ lines; the results they report are determined significantly by this choice of covering factors.  Since this paper is concerned with excitation temperatures, which are determined via the ratio of the two lines, properly accounting for the frequency dependence of the covering factor is vital to obtaining accurate results.

VLBI observations of \citet{mittal} show an extended source at low frequencies that splits image A into two point-like images separated at 15.4~GHz by $\sim$1.4~mas, corresponding to a physical separation in the lensing galaxy of 9.6~pc (for a flat universe with  H$_0=73$\kms~Mpc$^{-1}$, $\Omega_m=0.27$, and $\Omega_v=0.73$). The solid angle $\Omega_A$ 
follows a power law of index of $\alpha\approx-2.7$ (where $\Omega_A\propto\nu^\alpha$) as frequency increases from 1.65 to 15.65~GHz. (By comparing centimeter and millimeter observations, \citet{henkel05} argue that $-2<\alpha<-1$.) 
The relatively flat spectrum of the south-western source in image~A suggests that it is the flat-spectrum core of the quasar and the location of continuum emission seen in millimeter observations, where the  frequency-dependent jet would not be observable \citep{patnaik95}. Since the core is fully obscured ($f_\nu=1$) in millimeter observations, where the jet is not visible, the absorbing cloud lies along --- but is not necessarily centered upon --- the line of sight to it.  

At 8.6~GHz, where the jet and the core are not distinctly separated, the $\sim$15~mas$^2$ region containing the core emits $\sim$60\% of the flux of image A (\citet{patnaik95} measure 62(1)\% with VLBI observations at 15~GHz).  The total area of image A with flux density 5~times the rms background, found by extrapolating from the VLBI observations of \citet{mittal}, is $\sim$32~mas$^2$ at this frequency. A spherical molecular cloud of 30~pc diameter subtends $\sim$15~mas$^2$ ($\sim$47\% of $\Omega_A$ at 8.6~GHz) at $z=0.68$. Such a cloud gives a minimum covering factor of 0.3 if the core, which has $f_\nu=1$ at millimeter frequencies, is obscured only by a limb of the cloud; due to the comparable line widths observed at both centimeter and millimeter frequencies \citep{jethava07}, it is likely that the absorbing cloud is fortuitously aligned with image A and that the covering factor is close to 0.6 for a 30~pc cloud.  This is a strictly geometric argument, and weighting the covering factor by the fraction of flux that is obscured will give $f_{14.5}\geq0.6$ even for somewhat smaller clouds.  

The diameter of the cloud is likely to be significantly larger than 30~pc, since the optical obscuration of image A suggests the presence of large amounts of dust and possibly a giant molecular cloud \citep{grundahl95}; \citet{wik99} report a visual extinction of $A_v=850$ in front of image A.  Applying the relation between cloud size $S$ (equivalent to diameter for a spherical cloud) in pc and velocity dispersion $\sigma_v$ in\kms\ measured empirically with CO for Galactic clouds \citep{solomon87} of \begin{equation}\label{eqn:solomon}
  \frac{\sigma_v}{\rm km\ s^{-1}} = 1.0\pm0.1\left(\frac{D}{pc}\right)^{0.50\pm0.05},
\end{equation}
gives $S\approx225$~pc (CO line widths are 15\kms; \citealt{wik95}), suggesting a cloud much larger than the 650~mas$^2$ solid angle subtended by image A at 2.9~GHz.  While \citet{solomon87} measure the relation for clouds with $S<50$~pc and line widths $\lesssim8$\kms, this is a strong indication that the cloud is quite large, increasing the probability of a high covering factor. \citet{muller}, with high spectral resolution HCO$^+$ (2$-$1) observations, resolve the line into four $\sim$4.5\kms\ components, suggesting a clumpy medium of moderately sized clouds arrayed in front of image A, allowing the covering factor to remain constant or even increase with decreasing observation frequency (and increasing $\Omega_A$).  While the HCO$^+$ observations reveal that it is not a single large absorbing cloud, the presence of multiple clouds still allows for high covering factors. 

Agreement (1$\sigma$) between the observed optical depths and our LVG model (see Section \ref{sec:model}) requires $f_{4.8}>0.2$ for $f_{14.5}=0.6$ and $f_{4.8}>0.3$ for $f_{14.5}=1.0$. Matching optical depths requires $0.5<f_{4.8}<1.0$ for $f_{14.5}=0.6$ and $f_{4.8}=0.5$ for $f_{14.5}=1.0$.  Although likely, $f_{4.8}<f_{14.5}$ cannot be set $a\ priori$ due to the clumpy absorbing medium.  In subsequent analysis, we discuss results spanning the range of viable covering factors.

\subsection{Radiative Transfer Modeling}

\label{sec:model}

Formaldehyde is susceptible to non-LTE excitation in either direction (maser inversion or anti-inversion), so it is generally inappropriate to assume a single line excitation temperature for the molecule.  It is generally also not valid to attribute a single excitation  temperature to 
all of the centimeter $\Delta J=0$ lines, since all line excitation temperatures can differ. 
We employ the still valid assumption of statistical equilibrium in these H$_2$CO lines but allow all line excitation temperatures to float.

We used an LVG radiative transfer model to find relative level populations up to the 40th rotational energy state ($10_{37}$) of o-\form\ and the 41st ($7_{43}$) of p-\form.  Assuming statistical equilibrium, the model computes optical depths and excitation temperatures for a given formaldehyde density $n$(\form) and a hydrogen density \nh.  The model assumes an isothermal and constant density spherically symmetric cloud with large-scale turbulence or gravitational collapse producing large velocity gradients.  Following the observed velocity gradient of \citet{solomon87}, we set the velocity gradient to 1.0\kmspc.  \oNh\ values scale inversely with the gradient $dv/dr$ \citep{wang04}:
\begin{align}\label{eq:wang}
  \oNh &= [n({\rm o-}\form)/(dv/dr)]\times 3.08 \times 10^{18}\times\Delta \nu\notag\\&=[\nh\cdot X/(dv/dr)]\times 3.08 \times 10^{18}\times\Delta \nu,
\end{align}
where $\Delta \nu$ is the line width (FWHM), $n({\rm o-}\form)$ is the number density, and $X$ is the abundance of o-\form\ (not total \form) relative to H$_2$.
Since they are found via ratios of optical depths, which are proportional to \oNh, \nh\ and \tx\ results are independent of $dv/dr$. 
For a more thorough discussion of the LVG method, see \citet{sobolev} and \citet{gk74}.

\begin{figure*}[t]
\plottwo{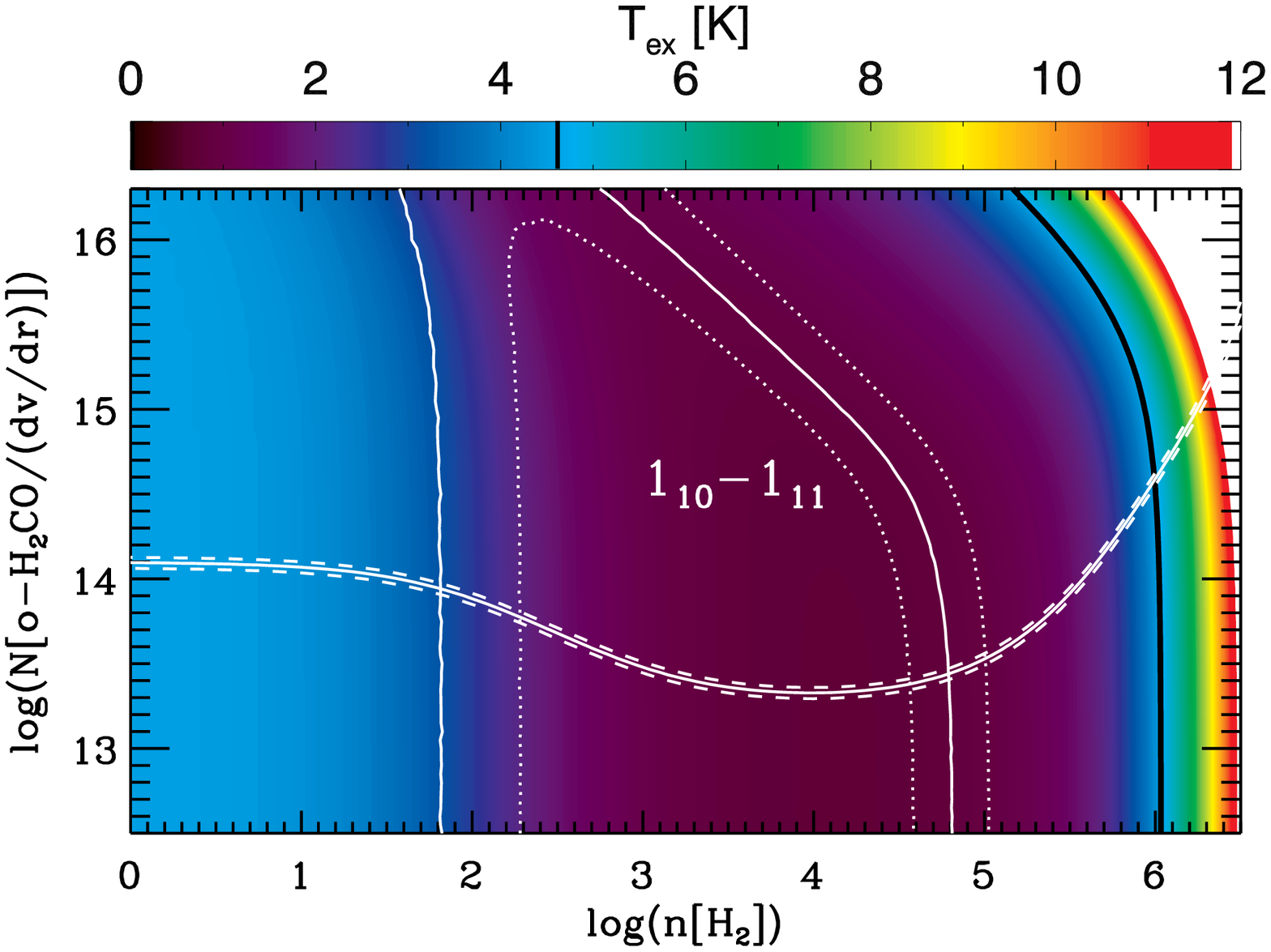}{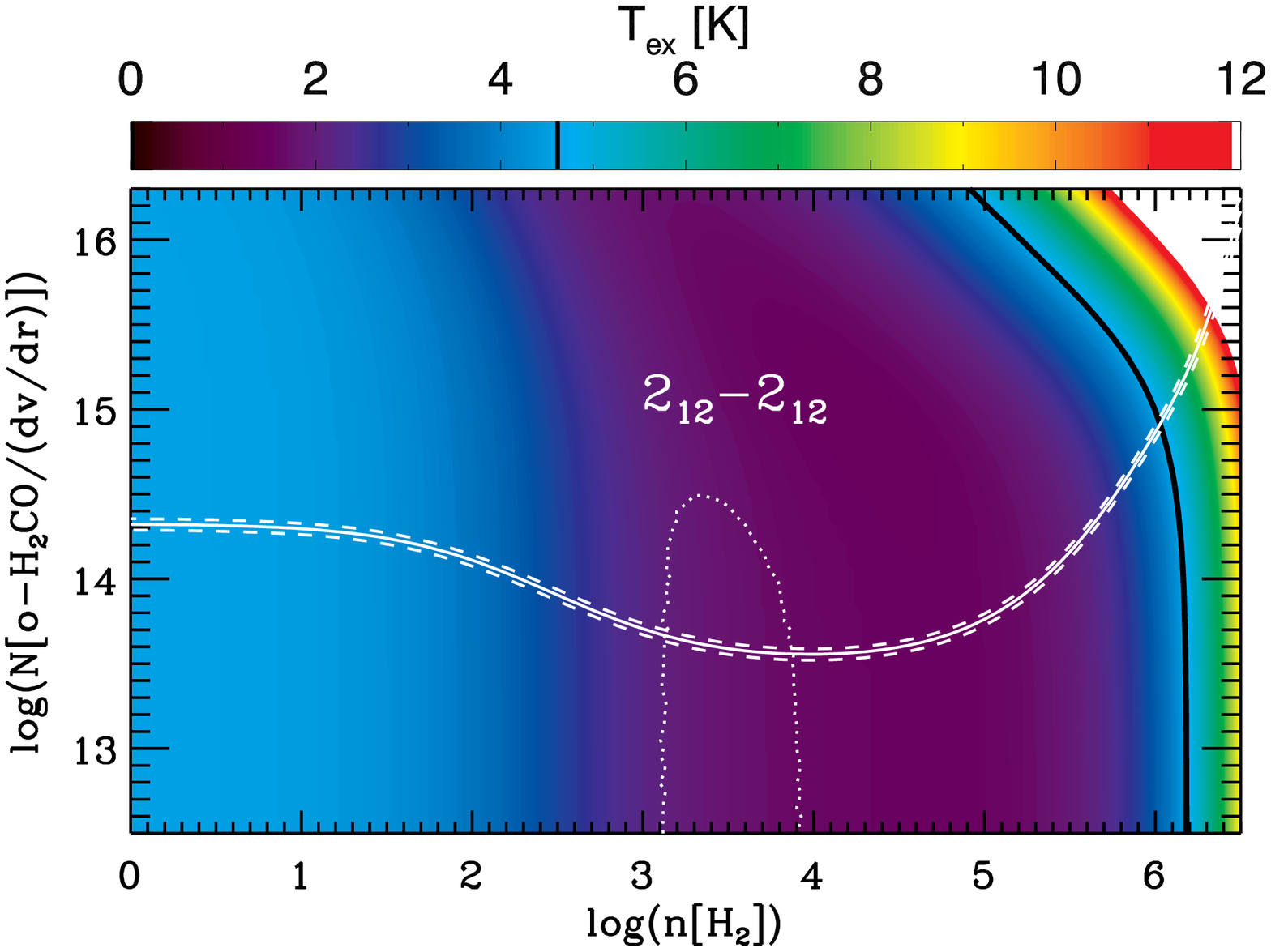}
\caption{
Contours trace excitation temperatures of the 
\cmone\ (left) and \cmtwo\ (right) transitions as a function of \nh\ and \oNh/\vgrad\ as calculated with
our LVG model.  Overplotted white lines represent the observed peak optical depth of the \cmone\ line ($\sim$horizontal) and the ratio of observed peak optical depths of the \cmone\ and \cmtwo\ lines ($\sim$vertical) with 68\% (1$\sigma$) confidence intervals. $T_{{\rm CMB}}=4.6$~K
at $z=0.68$, which we set as the background radiation temperature, and $\tkin=55$~K \citep{henkel05}.  The black line is the 4.6~K contour. CS observations and modeling firmly limit $\nh<2\times10^{4}$\pcc\ \citep{henkel05} --- the range of these plots exceed that limit to show the behavior of \form\ as a function of \nh\ --- and $T_{{\rm ex}}<T_{{\rm CMB}}$ for all points on the map below that limit. Column density ($y$-axis) is in units of\psc\ (\kmspc)$^{-1}$, and number density ($x$-axis) is in units of\pcc.  It is assumed throughout this paper that the velocity gradient $dv/dr$ is 1\kmspc.  The white at high \nh\ represents a ceiling to the contours at 12~K and not a physical plateau.  The upper right corner of the \cmone\ plot reaches $\sim$55~K, the kinetic temperature of the gas.  Overplotted white lines on the \cmone\ plot represent the solution if covering factors are $f_{4.8}=0.5$ and $f_{14.5}=0.8$; the central region is excluded, while a cloud with negligible collisions --- $\log(n[H_2])=0\pcc$ --- but unrealistic $X$ is not excluded by the ratio of optical depths at the 68\% confidence level if priors on $X$ are neglected.  The similar lines on the \cmtwo\ plot represent the constraints placed by observations if covering factors are $f_{4.8}=0.3$ and $f_{14.5}=1.0$; the central region is within the 68\% confidence region although there is no solution in the model that exactly matches the observed lines.  Differences in solutions shown on the two plots are due entirely to the selection of covering factors.
\label{fig:tx}}
\end{figure*}

Our model includes non-LTE collisional excitation of \form\ by H$_2$.  We use the He collisional cross sections calculated by \citet{green} as a proxy for the aspherical H$_2$; Green notes that the substitution of He for H$_2$ could cause errors up to 50\%\ in individual collisional excitation rates between levels, but that this is likely to translate to overall errors in level populations of order 20\%.  Using pressure-broadened H$_2$--\form\ and He--\form\ laboratory measurements of the millimeter transitions, \citet{mengel} find the He--\form\ collision rates to be ``very good,'' while the collisional cross sections of H$_2$--\form\ collisions are erroneous by up to a factor of 2 from the scaled collision rates ($\sqrt{\mu_{H_2}/\mu_{He}}$) with no apparent pattern to the errors.  

We evaluated the effect on excitation temperatures of such an error in collision rates by randomly altering collision rates with Gaussian variation with a mean of zero and a standard deviation of 50\%.  This overestimates the ``up to 50\%'' error quoted by \citet{green} but is close to that measured by \citet{mengel}.  For each of 25 sets of ``erroneous'' collisional rates, we generated the parameter space shown in Figure~\ref{fig:tx}.  The effect of the errors on line excitation temperatures is small at low densities, where collisions are infrequent (Section \ref{subsubsecColumn}); at moderate densities, the rms of the excitation temperatures of data sets is comparable to the mean excitation temperatures.  For the density measurements reported herein (see below), the mean excitation temperatures of our erroneous data sets are 0.85~K for the \cmone\ transition and 1.4--1.7~K for the \cmtwo\ transition; rms values are 1.0 and 1.0--1.5, respectively.
Taking an extreme scenario and uniformly doubling the collisional excitation rates of \citet{green} increases the effect of collisions at a given density, approximately halving the \nh\ values (0.3 dex) on the $x$-axis in Figure~\ref{fig:tx}.

The model used to determine excitation temperatures was generated with an ambient radiation temperature of  $T_{\rm CMB}=2.73(1+z)=4.60$~K and a kinetic temperature of 55~K. 
With a parameter space spanning 1\pcc\ $<\nh<10^7$\pcc\ and $3\times10^{12}$ \psc\ \vgrad\ $< \oNh/(dv/dr)<3\times10^{16}$\psc\ \vgrad\ 
with a sampling interval of 0.05 dex, we compared the observed peak optical depths to those generated with our LVG model. The model is weakly dependent on kinetic temperature: changes in modeled optical depth are less than observational error for $T_{K}=55\pm20$~K, and we present results only for $T_{\rm kin}=55$~K, as measured via NH$_3$ observations by \citet{henkel05}.

\begin{figure}
\epsscale{1.23}
\plotone{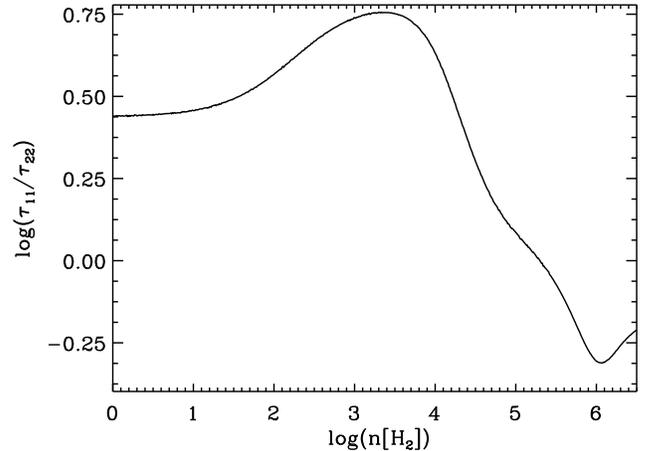}
\caption{Ratio of the optical depth of the \cmone\ transition to that of the \cmtwo\ transition as a function of \nh.  The modeled optical depths assume covering factors of unity and a constant \form\ abundance relative to H$_2$ of $X=\ten{-9}$.  The two-line densitometry method cannot produce unique results in regions where a single value of the optical depth ratio corresponds to two values of \nh\ unless the density is constrained by priors or ancillary information.  The non-unique region is enlarged if $X$ is not assumed to be constant, as can be seen in the two viable density regions plotted in the left image of Figure~\ref{fig:tx}, where the two possible densities are $\nh\approx6\times\ten{4}$\pcc\ (with $X\approx\ten{-11}$) and $\nh\approx60$\pcc\ (with $X\approx3\times\ten{-8}$).  If priors can discriminate between $\nh\lesssim2.5\times10^3$\pcc and $\nh\gtrsim2.5\times10^3$\pcc, solutions are unique and \form\ is a powerful densitometer.
\label{fig:ratio}}
\end{figure}

Of the four spectral components \citet{muller} identify in their high resolution HCO$^+$ spectrum, the strongest absorption occurs at the same velocity as both the \form\ \cmtwo\ line and the strongest component of the \cmone\ profile.  The ratio of the strongest \cmone\ absorption component and the \cmtwo\ line allows {\it in situ} densitometric measurements: the optical depth of one transition gives \oNh, and the ratio of two $K$-doublets gives \nh. 
Thus, with the two lines one can define a region of the modeled parameter space by matching observed optical depths to modeled ones.  When $\nh\lesssim2\times\ten{4}$, the ratio of lines can result in bifurcated solutions for \nh\ when using the two-line densitometry method employed herein --- one at moderate \nh\ and one at low \nh, as is shown in Figure~\ref{fig:ratio}. 

\begin{figure}[t]
\epsscale{1.23}
\plotone{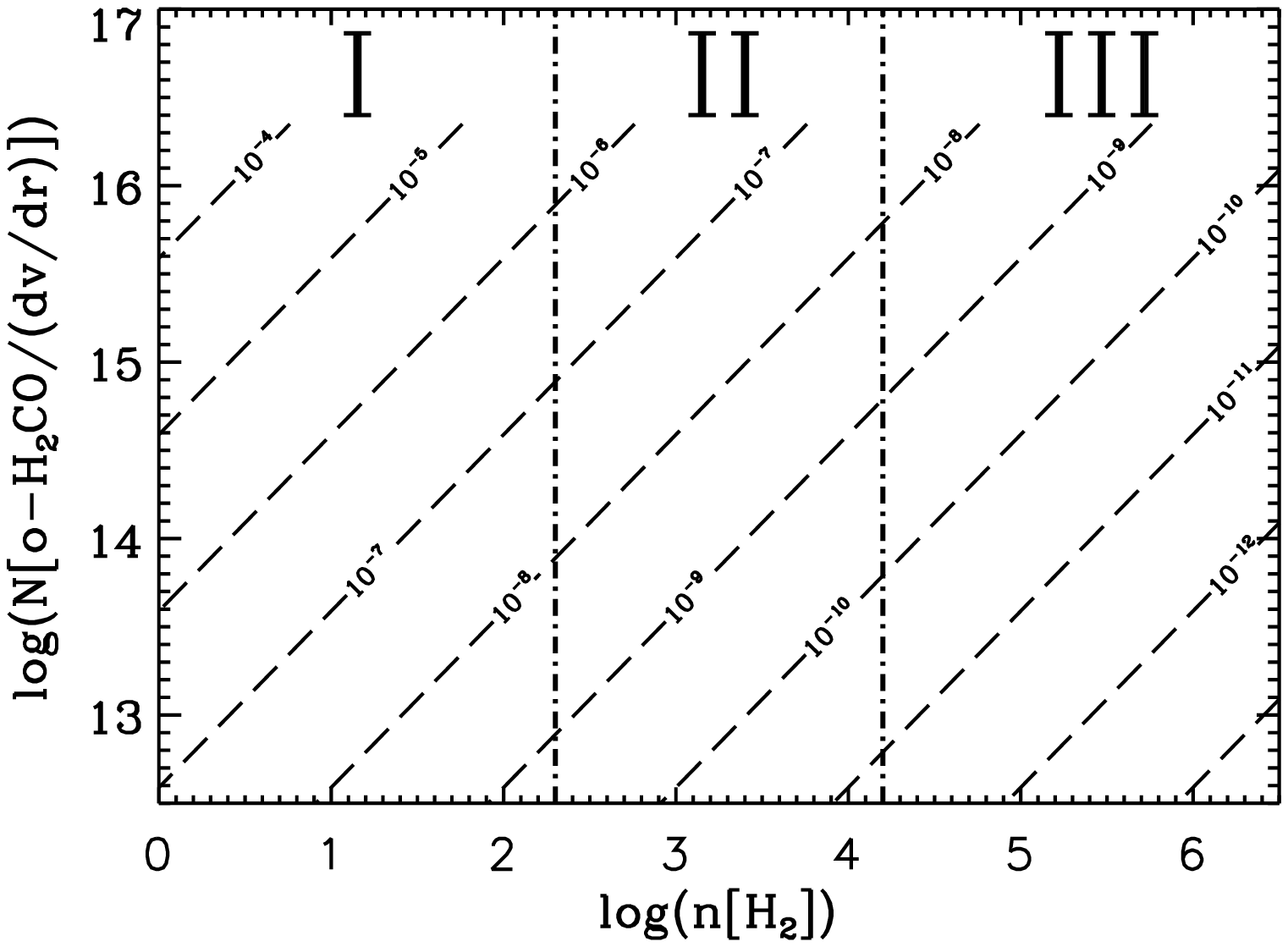}
\caption{Diagonal dashed lines mark the abundance  of o-\form\ relative to H$_2$ (see Equation~\ref{eq:wang}); vertical dot-dashed lines separate the regions, labeled at top, discussed in Section \ref{sec:model} and Table \ref{tab:regions}.  Typical Galactic abundances are of order $10^{-9}$, while measurements of \Nh\ \citep{jethava07} and $N(H_2)$ \citep{cw95,gerin97} place the total \form\ abundance between $2.5\times10^{-9}$ and $10^{-10}$ in \gal; the ortho/para ratio, which \citet{jethava07} measure as $\sim$2.8, drops the abundance of o-\form\ by $\sim$25\% relative to the total abundance of \form.
\label{fig:regions}}
\end{figure}

To break the degeneracy of solutions using observations of other molecules and abundance arguments, we divide the possible solutions into three regions separated at $\nh=2\times10^{2}$\pcc\ and $\nh=2\times10^{4}$\pcc, as 
shown in Figure~\ref{fig:regions} and 
described in Table~\ref{tab:regions}.  The Region I/II boundary at $\nh=2\times\ten{2}$ is an arbitrary division between low densities, where $\tx\approx\tcmb$ for both the \cmone\ and \cmtwo\ transitions, and moderate densities, where $\tx<\tcmb$.  The Region II/III boundary at $\nh=2\times10^{4}$\pcc\ is the firm upper limit placed by CS observations and modeling \citep{henkel05}.   Ranges of abundances in each region are calculated for all covering factors for which equilibrium model solutions exist, determining \oNh\ and \nh\ for all covering factors within these ranges, and converting column densities to number densities of o-\form\ with Equation~\ref{eq:wang}; maximum and minimum abundance values ($X$, Equation~\ref{eq:wang}) listed in Table \ref{tab:regions} are the maximum and minimum values of $n$(o-\form)/\nh\ in each region. Due to the high abundance $(\log[X]>-8.2)$, the low-density Region I is unlikely: typical Galactic abundances have a small dispersion around a ${\rm few} \times \ten{-9}$ (e.g., \citealt{evans75}; \citealt{dickel87}). Measurements of \Nh\ \citep{jethava07} and $N(H_2)$ \citep{cw95,gerin97} in \gal\ place the total \form\ abundance between $2.5\times10^{-9}$ and $10^{-10}$ in \gal; the ortho/para ratio, which \citet{jethava07} measure as 2.8 in their best model of the lines they observed, drops the abundance of o-\form\ by $\sim$25\%\ from the total \form\ abundance. These constraints agree with the value of $\nh\approx3.2\times10^{3}$\pcc\ estimated from NH$_3$ \citep[no confidence interval given]{henkel05}.

\begin{deluxetable}{ccccc}
\tablecolumns{5}
\tablewidth{0pc}
\tablecaption{Formaldehyde Radiative Transfer Model Regions: H$_2$ Number Density and \form\ Abundance
\label{tab:regions} }
\tablehead{
  \colhead{Region} &
  \colhead{Min. log(\nh)} &
  \colhead{Max. log(\nh)} &
  \colhead{Min. log($X$)} &
  \colhead{Max. log($X$)}
}
\startdata
I   & 0.0 & 2.3 &  $-$8.2 &  $-$5.6 \\
II  & 2.3 & 4.3 & $-$10.2 &  $-$8.2 \\
III & 4.3 & 5.5 & $-$10.3 & $-$10.2
\enddata
\tablecomments{Molecular hydrogen densities \nh, with units of\pcc, and o-\form\ abundances $X$ (Equation~\ref{eq:wang}) at the boundaries between the regions discussed in Section \ref{sec:model} and shown in Figure~\ref{fig:regions}.  The Region I/II boundary is an arbitrary division between low densities, where $\tx\approx\tcmb$, and moderate densities, where $\tx\ll\tcmb$.  The Region II/III boundary is the upper limit placed by CS observations and modeling \citet{henkel05}.  Ranges of abundances are calculated by determining \oNh\ and \nh\ for all covering factors with equilibrium model solutions within these ranges, and converting column densities to number densities of o-\form\ with Equation~\ref{eq:wang}; maximum and minimum $X$ values are the maximum and minimum values of $n$(\form)/\nh\ in each region. The typical \form\ abundance relative to H$_2$ in Galactic molecular clouds is $\sim$\ten{-9}, and the ortho/para ratio ranges from 1 to 3.  In \gal, the total \form\ abundance is $1-25\times\ten{-10}$, assuming an ortho/para ratio of 2.8.}
\end{deluxetable}

Fitting observed optical depths to our LVG model within the above constraints and forcing $f_{14.5}>0.6$ gives $2\times10^{3}\pcc <\nh<1\times\ten{4}\pcc$ (1$\sigma$) for the covering factors discussed above.  Agreement between observations and modeling requires $f_{4.8}/f_{14.5}\gtrsim1/3$, while the ratio of solid angles in image A is $\Omega_{4.8}/\Omega_{14.5}\approx20$. Since the covering factors are poorly constrained, results for \nh\ and \tx\ cannot have proper confidence intervals.  However, our reported ranges include all reasonable covering factors ($f_{14.5}>0.6$, $f_{4.8}/f_{14.5}\gtrsim1/3$) and are thus robust.  We list \nh\ and \tx\ for various covering factors in Tables \ref{tab:density}$-$\ref{tab:tx2}; $\tx<\tcmb$ for all physical pairs of covering factors. 

\subsection{Column Density and Excitation Temperatures}
\label{subsubsecColumn}
We find   $2.5\times10^{13}\psc<\oNh<8.9\times10^{13}\psc$ (assuming a velocity gradient of 1\kmspc), with the range determined by the viable covering factors ($f_{14.5}>0.6$, $f_{4.8}/f_{14.5}\gtrsim1/3$) and constraints on molecular hydrogen densities ($2\times\ten{3}\pcc<\nh<1\times\ten{4}$ \pcc). Since these results cover all reasonable covering factors, the limits are hard and conservative.  The total \form\ column density, assuming $\op=2.8$ from the best models of \citet{jethava07}, which is determined primarily from millimeter observations where covering factors are unity and not a concern, is $3.4\times\ten{13}$---$1.2\times\ten{14}$ \psc. The column density scales inversely with the assumed velocity gradient.
\begin{figure}[t]
\epsscale{1.23}
\plotone{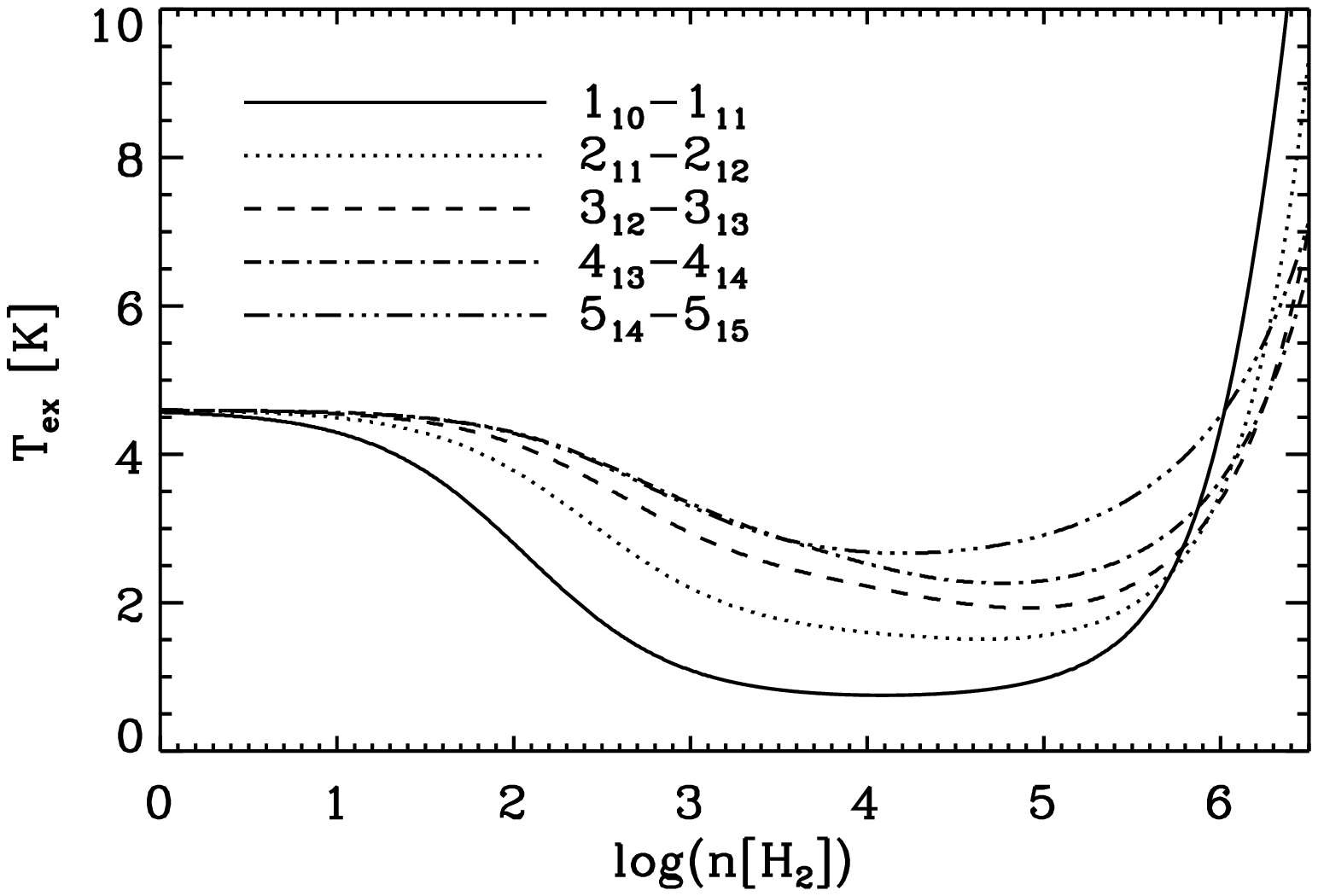}
\caption{
Excitation temperatures of the lowest five $K$-doublet transitions as a function of \nh\ for $\oNh=\ten{14}$\psc, as calculated with our LVG model. This figure is a horizontal slice through Figure~\ref{fig:tx} with additional $K$-doublet lines added.  The divergence of the $K$-doublet excitation temperatures as a function of \nh\ makes the ratio of the any pair of lines a sensitive densitometer.  However, since the ratios converge to unity at low \nh\ (as \tx\ approaches \tcmb) and cross it again at $\nh\approx\ten{6}$ (as all excitation temperatures go to \tkin\ at high densities), the ratios cannot provide a unique global solution unless the covering factor is known {\it a priori} or other information, such as observations of other molecules, is available to break the degeneracy (Section \ref{sec:model}).\label{fig:u}
}
\end{figure}

\begin{deluxetable}{cccc}
\tablecolumns{4}
\tablewidth{0pc}
\tablecaption{Model-Derived Number and Column Densities For Various Covering Factors}
\tablehead{
  \colhead{$f_{4.8}$} &
  \colhead{$f_{14.5}$} &
  \colhead{log$(n({H_2}))$} &
  \colhead{log$(N({o-H_2CO})/(dv/dr))$}\\
  \colhead{}&
  \colhead{}&
  \colhead{(cm$^{-3}$)}&
  \colhead{(cm$^{-2}$ (km s$^{-1}$ pc$^{-1}$)$^{-1}$)}
}
\startdata\vspace{+1mm}
 0.50& 0.60 & 5.05$^{+ 0.15}_{ -0.15}$ & 13.55$^{+ 0.10}_{ -0.10}$ \\\vspace{+1mm}
 & & 0.03$^{+ 1.47}_{ -0.03}$ & 14.10$^{+ 0.00}_{ -0.10}$ \\\vspace{+1mm}
 0.50& 0.80 & 4.80$^{+ 0.20}_{ -0.20}$ & 13.45$^{+ 0.10}_{ -0.05}$ \\\vspace{+1mm}
 & & 1.80$^{+ 0.45}_{ -1.80}$ & 13.95$^{+ 0.15}_{ -0.15}$ \\\vspace{+1mm}
0.30 & 1.00 & 3.53$^{+0.37}_{-0.38}$ & 13.60$^{+ 0.05}_{ -0.05}$\\\vspace{+1mm}
 1.00& 1.00 & 5.15$^{+ 0.15}_{ -0.10}$ & 13.30$^{+ 0.10}_{ -0.05}$\vspace{+1mm}
\enddata
\tablecomments{Confidence intervals are set at 1$\sigma$ (68\%) but are non-Gaussian.  Where more than one solution exists within 1$\sigma$ observational error, all solutions are given. An error of `0.00' indicates that the modeled density did not change significantly across the region of density space defined by fitting observed optical depths to those derived with our LVG model. Number densities above $2\times\ten{4}$\pcc\ contradict the firm upper limit obtained from LVG models of CS observations \citep{henkel05}.\label{tab:density}}
\end{deluxetable}

\begin{deluxetable*}{@{\extracolsep{-.075in}}cccccccccc}
\tablecolumns{10}
\tablewidth{0pc}
\tablecaption{Modeled $K$-doublet ($\Delta J=0$) o-\form\ Line Excitation Temperatures}
\tablehead{
  \colhead{$f_{4.8}$} &
  \colhead{$f_{14.5}$} &
  \colhead{log(\nh)}&
  \colhead{$1_{10}$$-$$1_{11}$} &
  \colhead{$2_{11}$$-$$2_{12}$} &
  \colhead{$3_{12}$$-$$3_{13}$} &
  \colhead{$4_{13}$$-$$4_{14}$} &
  \colhead{$5_{14}$$-$$5_{15}$} &
  \colhead{$6_{15}$$-$$6_{16}$} &
  \colhead{$7_{16}$$-$$7_{17}$}
\\
\colhead{} &
\colhead{} &
\colhead{(cm$^{-3}$)} &
\colhead{4.8~GHz} &
\colhead{14.5~GHz} &
\colhead{29.0~GHz} &
\colhead{48.3~GHz} &
\colhead{72.4~GHz} &
\colhead{101.3~GHz}&
\colhead{135.0~GHz}
}
\startdata\vspace{+1mm}
0.50 & 0.60 & 5.05$^{+0.15}_{-0.15}$&0.97$^{+3.26}_{-0.09}$ & 1.57$^{+2.90}_{-0.03}$ &1.98$^{+2.55}_{-0.01}$&2.34$^{+2.22}_{-0.04}$&2.94$^{+1.61}_{-0.11}$&3.63$^{+0.76}_{-0.14}$&{\bf 4.77}$^{+0.20}_{-0.79}$\\\vspace{+1mm}

 &  & 0.03$^{+1.47}_{-0.03}$& 4.57$^{+0.00}_{-1.23}$ & 4.59$^{+0.00}_{-0.51}$ &4.59$^{+0.00}_{-0.28}$&{\bf 4.60$^{+0.00}_{-0.19}$}&{\bf 4.60$^{+0.00}_{-0.20}$}&4.58$^{+0.00}_{-0.65}$&4.41$^{+0.00}_{-0.48}$\\\vspace{+1mm}
0.50 & 0.80 & 4.80$^{+0.20}_{-0.20}$& 0.84$^{+3.42}_{-0.06}$ & 1.53$^{+2.95}_{-0.00}$ &1.98$^{+2.57}_{-0.01}$&2.29$^{+2.27}_{-0.00}$&2.79$^{+1.77}_{-0.07}$&3.43$^{+1.00}_{-0.12}$&{\bf 4.53$^{+0.20}_{-0.53}$}\\\vspace{+1mm}
 &  & 1.80$^{+0.45}_{-1.80}$& 3.24  $^{+1.33  }_{-0.99}$ & 4.03 $^{+0.56}_{-0.63}$ &4.30$^{+0.29}_{-0.43}$&{\bf 4.40$^{+0.20}_{-0.30}$}&{\bf 4.39$^{+0.21}_{-0.31}$}&3.92$^{+0.66}_{-0.44}$&3.93$^{+0.48}_{-0.03}$\\\vspace{+1mm}
0.30 & 1.00 & 3.53$^{+0.37}_{-0.38}$ & 0.80$^{+3.34}_{-0.05}$ & 1.75$^{+2.68}_{-0.10}$ & 2.51$^{+2.01}_{-0.20}$ & 2.89$^{+1.66}_{-0.27}$ & 2.87$^{+1.67}_{-0.16}$&3.06$^{+1.29}_{-0.00}$&3.93$^{+0.07}_{-0.04}$\\\vspace{+1mm}
1.00 & 1.00 &5.15$^{+0.15}_{-0.10}$& 1.04$^{+3.31}_{-0.08}$ & 1.60$^{+2.91}_{-0.03}$ &2.01$^{+2.55}_{-0.03}$&2.39$^{+2.18}_{-0.05}$&3.03$^{+1.54}_{-0.10}$&3.76$^{+0.71}_{-0.14}$&{\bf 4.92}$^{+0.25}_{-0.90}$\vspace{+1mm}
\enddata
\tablecomments{\label{tab:tx}Excitation temperatures for selected pairs of covering factors $f_{4.8}$ and $f_{14.5}$; boldfaced temperatures have $\tx\geq\tcmb$ ($\tcmb=4.60$~K) within the listed 68\% (non-Gaussian) confidence intervals.  Frequencies are rest frequencies, and all excitation temperatures have units of K. Regions listed with errors of `0.00' do not change excitation temperature significantly across the region of density space defined by fitting observed optical depths to those derived with our LVG model. Multiple excitation temperatures for a given pair of covering factors correspond to solutions in Table \ref{tab:density} where a pair of optical depths did not produce a unique solution.  Low density solutions ($\nh<\ten{2}$\pcc) have $\tx=\tcmb\ (z=0.68466) = 4.60$~K because collisions are insufficient to be a significant excitation mechanism. Of the $K$-doublet transitions, only the \cmone\ and \cmtwo\ lines have been observed in \gal\ (see Table \ref{tab:H2CO}).
}
\end{deluxetable*}

\begin{deluxetable*}{cccccccccc}
\tablecolumns{10}
\tablewidth{0pc}
\tablecaption{Modeled $\Delta J=1$ \form\ Line Excitation Temperatures}
\tablehead{
{$f_{4.8}$} &{$f_{14.5}$} &   \colhead{log(\nh)}&
{$2_{12}$$-$$1_{11}$ (O)} & {$2_{11}$$-$$1_{10}$ (O)} & {$1_{01}$$-$$0_{00}$ (P)} & {$2_{02}$$-$$1_{01}$ (P)}\\
&&\colhead{(cm$^{-3}$)} &
\colhead{140.8~GHz} &
\colhead{150.5~GHz} &
\colhead{72.8~GHz} &
\colhead{145.6~GHz}}
\startdata\vspace{+1mm}
0.50 & 0.60 &5.05$^{+0.15}_{-0.15}$&7.73$^{+1.45}_{-3.13}$ & 6.70$^{+0.89}_{-2.10}$ & 17.40$^{+14.90}_{-12.80}$ & 6.74$^{+0.84}_{-2.14} $ \\ \vspace{+1mm}
 &  & 0.03$^{+1.47}_{-0.03}$&4.60$^{+0.00}_{-0.00}$ & 4.60$^{+0.00}_{-0.00}$ & 4.60$^{+0.00}_{-0.00}$ & 4.60$^{+0.00}_{-0.00} $ \\\vspace{+1mm}
0.50 & 0.80 & 4.80$^{+0.20}_{-0.20}$& 6.31$^{+1.06}_{-1.71}$ & 5.78$^{+0.69}_{-1.18}$ & 9.77$^{+5.23}_{-5.17}$ & 5.84$^{+0.67}_{-1.24} $ \\\vspace{+1mm}
&& 1.80$^{+0.45}_{-1.80}$&4.60$^{+0.00}_{-0.00}$ & 4.60$^{+0.00}_{-0.00}$ & 4.60$^{+0.01}_{-0.00}$ & 4.60$^{+0.00}_{-0.00} $ \\\vspace{+1mm}
0.30 & 1.00 & 3.53$^{+0.37}_{-0.38}$ & 4.73$^{+0.06}_{-0.13}$ & 4.68$^{+0.05}_{-0.08}$ & 4.88$^{+0.16}_{-0.28}$ & 4.70$^{+0.05}_{-0.10}$\\\vspace{+1mm}
1.00 & 1.00&5.15$^{+0.15}_{-0.10}$& 8.63$^{+1.97}_{-4.03}$ & 7.26$^{+1.14}_{-2.66}$ & 25.20$^{+42.70}_{-20.60}$ & 7.27$^{+1.08}_{-2.67} $\vspace{+1mm}
\enddata
\tablecomments{\label{tab:tx2}Similar to Table \ref{tab:tx} but for $\Delta J=1$ transitions of both o-\form\ (O) and p-\form\ (P).  All excitation temperatures have units of K.  All transitions in this table have been observed in \gal\ \citep{jethava07}.  Low density solutions ($\nh<\ten{2}$\pcc) have $\tx=\tcmb\ (z=0.68466) = 4.60$~K because collisions are insufficient to be a significant excitation mechanism.
}
\end{deluxetable*}

Regardless of covering factors (although tables herein only list several, the entire range of reasonable covering factors was evaluated), Figure~\ref{fig:tx} shows that the excitation temperatures of both observed lines are below \tcmb\ within the constraints placed by CS, which require $\nh\lesssim2\times\ten{4}$ \pcc. Figures~\ref{fig:tx} and \ref{fig:u} show clearly that, at $z=0.68$, the \form\ centimeter lines will be anti-inverted. 
The non-LTE excitation is density dependent:
\begin{enumerate}
  \item $\nh\lesssim\ten{2} \pcc$: \form\ excitation is dominated by CMB photons.  All excitation temperatures are held near to the microwave background and the system is nearly in radiative equilibrium with the CMB because collisions with H$_2$ are too infrequent to pump the \form\ $K$-doublet population into anti-inversion.
  \item $\ten{2}\pcc\lesssim\nh\lesssim3\times\ten{5}\pcc$: Collisions with H$_2$ anti-invert the \form\ $K$-doublet population. The millimeter transitions, however, have $\tx\geq\tcmb$.
  \item $\nh\gtrsim3\times\ten{5}\pcc$: High densities increase collision rates and drive \form\ toward thermal equilibrium with H$_2$, with a kinetic temperature of 55~K.
\end{enumerate}

Figure~\ref{fig:u}, a horizontal slice through Figure~\ref{fig:tx}, shows a `U' shape of excitation temperatures.  The two possible values of \tx\ and \nh\ for a given ratio of observed optical depths cause the degeneracy discussed in Section \ref{sec:model} and shown in Figure~\ref{fig:ratio}. The \nh-dependence of centimeter excitation temperatures makes \form\ a sensitive {\it in situ} densitometer if the degeneracy can be broken with {\it a priori} knowledge of physical conditions such as \form\ abundance or limits on \nh.  

\section{Discussion}\label{sec:discussion}
\subsection{Comparison to Excitation in Galactic Molecular Clouds}

Galactic dark clouds show a decrement in the excitation temperature of the \cmone\ line relative to the CMB of $\tx-T_{\rm CMB}\simeq-0.7$ to $-0.4$ K \citep{evans75}.
An LVG radiative transport model applied to Galactic giant molecular clouds derives comparable decrements, ranging from $-$1 to 0~K \citep{henkel80}.  The cloud studied in this paper shows a much larger decrement, with 
$\tx-T_{\rm CMB}\lesssim-3.6$~K (for $2\times\ten{3}\pcc<\nh<1\times\ten{4}\pcc$). For the \cmtwo\ transition, the decrement is $\sim-$2.6 to $-$3.1~K.  The magnified decrement is due to the collisional dependence of excitation temperatures (Section \ref{subsubsecColumn}): CMB excitation is only significant in diffuse clouds with $\nh\lesssim\ten{2}$\pcc, so for higher density clouds the excitation temperature is nearly constant regardless of the CMB temperature.  Increasing the background temperature thus magnifies the $\tx-\tcmb$ decrement  proportionally to $(1+z)$.

Because $\tx-\tcmb<0$ for both transitions, 
the cloud would be observable in absorption against the CMB in both lines without the background quasar -- if the cloud were in a different place in the lensing galaxy rather than aligned with image~A or in a galaxy with no background continuum source, for instance.  In the case where the cloud is matched to the telescope beam (no beam dilution) at $z=0.68$ and the covering factor is unity, the observed \cmone\ line temperature would be 
\begin{equation}T_{{\rm line}}=(1-e^{-\tau})(\tx-\tcmb)/(1+z)\approx-36{\rm \ mK}.\end{equation}
For a non-unity covering factor and small optical depth, as seen in \gal, $T_{{\rm line}}$ is approximately inversely proportional to the covering factor.  Such a cloud illuminated by the CMB will be observable with future instruments such as the EVLA.  Since \tx\ is small and nearly constant with redshift and \tcmb\ scales as $1+z$, this raises the possibility of surveying molecular gas in galaxies {\it regardless of distance} or chance alignment with sources of background illumination. 

Models indicate that the anti-inversion persists in the $K_a=1$ $K$-doublets at least up to the $6_{15}$$-$$6_{16}$ o-\form\ transition for all reasonable covering factors (which determine the measured molecular hydrogen density), and for some covering factors the anti-inversion persists into the $7_{16}$$-$$7_{17}$ transition (Table \ref{tab:tx}). 
Although the highest energy lines may be difficult to detect, 
all of the anti-inverted transitions should in principle be observable in absorption against the CMB. 

$K$-doublet transitions with $K_a=2$ (p-\form) and $K_a=3$ (o-\form) are even more strongly anti-inverted than those in the $K_a=1$ ladder, but relative to the $K_a=1$ $K$-doublets the higher energy of the $K_a>1$ states (the lowest $K_a=2$ $K$-doublet level, $2_{21}$ at 57.6~K, has approximately the same energy as the $5_{15}$ level at 62.5~K) generally makes them more optically thin and difficult to observe given the low temperatures of molecular clouds.  Our models indicate that the $2_{20}-2_{21}$ transition (71.1~MHz rest frequency), the lowest $K$-doublet of the $K_a=2$ ladder, has an excitation temperature less than 10~mK in \gal. Due to the relatively warm temperature of the gas observed in \gal\ (55~K), the optical depth of the transition is predicted to be lower than that of the observed \cmone\ o-\form\ line by only a factor of $\sim$5 and higher by a factor of $\sim$20 than would be expected in an otherwise equivalent cloud with a kinetic temperature of 15~K (assuming an ortho/para ratio of 2.8; \citealt{jethava07}).  The next lowest $K$-doublet of the $K_a=2$ ladder, the $3_{21}$$-$$3_{22}$ transition (355.6~MHz rest), is predicted to have an optical depth that is an order of magnitude lower than that of the $2_{20}$$-$$2_{21}$ line. All other $K_a>1$ $K$-doublets are weaker by yet another order of magnitude or more 
than the $3_{21}$$-$$3_{22}$ line. In the hypothetical case of a similar gas-rich galaxy illuminated only by the CMB at $z=0.68$ (the flux from the quasar illuminating \gal\ is not well known at meter wavelengths), the observed line temperatures for the lowest $K_a=2$ $K$-doublets are 
predicted to be between  $-$4 and $-$11 mK for the $2_{20}$$-$$2_{21}$ line and $-$0.3 and $-$1~mK for the $3_{21}$$-$$3_{22}$ line (assuming $\Nh=5\times\ten{13}$\psc\ and an ortho/para ratio of 2.8, with the range set by our measured range of values for the molecular hydrogen density in the cloud). While these low-frequency lines will be dwarfed by the foreground Galactic continuum and difficult to observe for their own sake, this modeling indicates that \form\ $K_a=2$ absorption lines in gas-rich galaxies present potential contaminants to Epoch of Reionization studies.   



\subsection{Comparison to Previous Results}
\citet{jethava07} measure $\oNh=3.8\times10^{13}$\psc\ and $\nh<\ten{3}$\pcc\ (with a best fit of $\nh=200$\pcc) with their non-LTE LVG analysis of six \form\ lines in \gal. \citet{menten96}, assuming LTE, measure $\oNh=1.2\times10^{13}$\psc\ from their observation the \cmtwo\ line, although they estimate that anti-inversion of the $K$-doublet would reduce their result by 50\%, well below the column densities reported by \citet{jethava07} and our range of $2.5\times10^{13}-8.9\times10^{13}\psc$ (for $dv/dr=1\kmspc$). 

Our \cmone\ spectrum has a significantly higher signal to noise ratio and better spectral resolution than previous observations, and reliance solely on it and our new \cmtwo\ data to determine densities precludes high-frequency atmospheric variability while magnifying the effect of uncertain covering factors. Since the centimeter $K$-doublet lines measure \nh\ with greater precision than the millimeter lines can, the trade-off makes the \citet{jethava07} result less accurate for determining \nh\ and \tx\ (which they do not address) but more reliable for measuring \oNh, since they rely heavily on nearly thermal millimeter transitions which have covering factors of unity.  The optimal combination of the two methods uses the centimeter lines to constrain \nh\ (and \tx, since it is closely related) and the millimeter lines to constrain \oNh.  \tx\ and \nh\ are nearly independent of \oNh\ at the column densities observed in this source (Figure~\ref{fig:tx}).



Our grid-search method is a more comprehensive analysis of the \nh-\oNh\ density space than that used by \citet{jethava07}, who compared their observations only to LVG models with $\nh=0$, 200, 500, 1000, and 3000\pcc\ to determine a best-fit ($\chi^2$) value.  The small values of \nh\ sampled by \citet{jethava07} also neglect the high-\nh\ solutions discussed in Section \ref{sec:model}, essentially assuming their conclusion that the absorbing cloud is diffuse and low-density and not considering the alternative higher density regime.


\section{Conclusions}\label{sec:conclusions}

With new measurements of the \cmone\ and \cmtwo\ transitions of \form\ in \gal, we have measured the molecular hydrogen number density and \form\ column density.  Using an LVG model of the physical conditions of the absorbing cloud in the gravitational lens and assuming a velocity gradient of 1\kmspc, we measure \oNh\ to be in the range $2.5\times10^{13}-8.9\times10^{13}\psc$, 
where the uncertainty in values is caused by the uncertainty in covering factors (in the optically thin limit, the column density is inversely proportional to the covering factor). \citet{jethava07} measure an ortho/para ratio of $2.0-3.0$ (with a best value of 2.8) and a second component to the profile that adds $\sim$15\% to the total column density; these two contributions increase the total column \Nh\ by $\sim$55\%\ above our measurement of \oNh\ in the primary absorption component.  Since collisions with hydrogen are the dominant excitation mechanism, we are also able to measure $\nh= 2\times10^{3}-1\times\ten{4}\pcc$, which again spans the range of viable covering factors.  

Modeled excitation temperatures show both observed centimeter lines to be anti-inverted relative to the 4.6~K CMB at $z=0.68$, with $\tx(\cmone)\approx1$~K and $\tx(\cmtwo)\lesssim2$~K.  
 Excitation temperatures of the millimeter lines are $\geq$4.60~K, as expected.  These results show that the dominant absorbing cloud in the \gal\ lens is analogous to Galactic molecular clouds that show anti-inversion of the $K$-doublet transitions, and the centimeter \form\ lines could be observed {\it in absorption against the CMB} if the background source was removed or if the cloud was viewed along a different line of sight. 

Since \tx\ is determined by collisions in dense clouds, the source-frame decrement $\tx-\tcmb$ is magnified when \tcmb\ is increased (n.b., at high redshift).  The anti-inverted centimeter transitions thus could provide absorption lines against the CMB that would trace molecular gas {\it independently of distance} when observed with a telescope capable of resolving the star-forming regions in galaxies.  Future work targeting additional objects will quantify the change in $\tx-\tcmb$ with redshift and test these results.




\acknowledgments
We are indebted to C. Henkel for providing the LVG code used in this analysis and to the anonymous referee who suggested that we model the excitation temperatures of $K_a>1$ $K$-doublets and checked our calculations with an independent radiative transfer code.

Support for this work was provided by NASA through Hubble Fellowship grant
no. HST-HF-01183.01-A awarded by the 
Space Telescope Science Institute, which is 
operated by the Association of Universities for Research in Astronomy, 
Incorporated, under NASA contract NAS5-26555.  We also acknowledge the support of the NSF through award GSSP07-0015 from the NRAO and NSF grant AST-0707713.

The LUNAR consortium (\url{http://lunar.colorado.edu}), headquartered at
the University of Colorado, is funded by the NASA Lunar Science
Institute (via Cooperative Agreement NNA09DB30A) to investigate
concepts for astrophysical observatories on the Moon.

This research has made use of the NASA/IPAC Extragalactic Database (NED) 
which is operated by the Jet Propulsion Laboratory, California Institute 
of Technology, under contract with the National Aeronautics and Space Administration.




\end{document}